\documentclass{article}

\input pz.sty
\usepackage{graphicx}

\begin{document}

\PZhead{1}{40}{2020}{28 February}

\PZtitletl{Photometric observations of two type Ic-BL}{Supernovae: 
2016\lowercase{coi} and 2018\lowercase{ebt}}

\PZauth{D. Yu. Tsvetkov, N. N. Pavlyuk, V.A. Echeistov}
\PZinst{Sternberg Astronomical Institute, University Ave.13,
119992 Moscow, Russia; e-mail: tsvetkov@sai.msu.su}

\begin{abstract}
CCD $BVRI$ photometry is presented for the type Ic-BL Supernovae
2016coi and 2018ebt. The shape of the light curves for both 
objects is typical for this class of SNe. 
A change of brightness decline rate at about 100 days
after maximum light is observed for SN 2016coi. The light curves of 
SN 2018ebt are best matched
by those of SN Ic-BL 2002ap. For SN 2018ebt, we derived the dates and 
magnitudes
of maximum light, rates of decline at late stages of evolution and 
presented evidence for extinction in the host galaxy
being negligible. 
\end{abstract}

\bigskip
\PZsubtitle{\bf Introduction}

A large fraction of massive stars end their lives with  
explosions due to the 
gravitational collapse of their cores, such events are 
recognized as core-collapse supernovae (CCSNe). 
Among the CCSNe 
types Ib, Ic and IIb represent 
the "stripped-envelope" (SE) 
sub-category where the progenitor's outer envelope of hydrogen
and/or helium is partially or completely removed before the
explosion (Filippenko, 1997). 
A sub-population of Type Ic (He poor or absent) SNe are characterized
by very broad absorption lines in their 
spectra, which results from high expansion velocities of the 
ejecta. These events are  
designated as "broad-line" Ic SNe (Ic-BL) (Valenti et al. 2008). 
A small fraction of 
them are linked with long-duration gamma ray bursts (GRBs), 
e.g. SN~1998bw/GRB~980425 (Galama et al. 1998), while no such  
association is observed in a majority of the Ic-BL events.

In this paper, we present the results of photometric observations
of two recent bright type Ic-BL SNe: SN 2016coi and SN 2018ebt.

SN 2016coi was discovered on 2016-05-27.55 UT 
at the $V$-band magnitude $\sim 15.7$
(Holoien et al. 2016)
by the All Sky 
Automated Survey for Supernovae (ASAS-SN)\footnote{http://www.astronomy.
ohio-state.edu/~assassin/index.shtml};
it was located in the Sm galaxy UGC 11868, 
$z=0.0036$, at 
$\alpha=21\hr59\mm04\sec.14, \delta=+18\deg 11\arcm 10\arcs.46$ (J2000). 
The SN was offset by 31\arcs.7 north and 7\arcs.9 west from 
the center of the host galaxy. 
It was classified as 
a pre-maximum "broad lined" Type Ic SN on 2016-05-28.52 UT
(Elias-Rosa et al. 2016).
SN 2016coi was extensively studied by Yamanaka et al. (2017),
Prentice et al. (2018), Kumar et al. (2018), and  Terreran et al. (2019).   
 
SN 2018ebt was discovered by ATLAS\footnote{http://fallingstar.com/home.php}
on 2018-07-21.49 UT at magnitude
15.58 in the 'orange' filter. The object was fainter than 19.29 mag
in the same filter on 2018-07-19.50 UT. The SN was located at 
$\alpha=20\hr41\mm54\sec.99,  \delta=+64\deg 12\arcm 52\arcs.6$
(J2000.0)
The discovery was reported by the Transient Name Server 
(TNS)\footnote{https://wis-tns.weizmann.ac.il/object/2018ebt}. 
The host galaxy is GALEXASC J204152.80+641238.5, and the offsets from 
the center are $14\arcs.7$ east and $8\arcs.7$ north.

Moran et al. (2018) reported that the spectroscopic observation had been 
performed with the 2.56-m Nordic Optical 
Telescope equipped with ALFOSC 
on 2018-07-24.06 UT. The closest matches were found with SNe Ic-BL 
with phases at a few days before the peak or around the peak, and the redshifts
between 0.02-0.03 or less. TNS also reported that 
SN 2018ebt had been classified by Masayuki Yamanaka 
as an SN Ib/c at $z$=0.005 using the 
spectrum taken by 1.5-m Kanata telescope on 2018-07-23.58 UT.
Dugas et al. (2018) reported that spectrum obtained on 2018-08-14.86 UT
with the Palomar 60-inch telescope had indicated SN type Ic and $z$=0.01.

\bigskip
\bigskip
\PZsubtitle{\bf Observations and reductions}

We carried out photometric observations of 
SN 2016coi from 2016-08-29 to 2017-01-30 and of
SN 2018ebt from 2018-07-31 to
2018-11-25 using mainly the 60-cm reflector of 
Crimean Observatory of Sternberg Astronomical Institute (SAI) (C60) and
the 70-cm SAI
reflector in Moscow (M70), some images were also obtained at 
the 1-m telescope of Simeiz Observatory (S100), the 50/70-cm meniscus 
telescope of 
SAI Crimean Observatory (C50), and the 60-cm reflector of 
Star\'a Lesn\'a Observatory 
in Slovakia (S60).

The standard image reductions and photometry were made using 
IRAF\footnote{IRAF is distributed by the National Optical Astronomy Observatory,
which is operated by AURA under cooperative agreement with the
National Science Foundation}.
Photometric measurements of SNe were made relative to 
local standard stars using PSF fitting with the IRAF DAOPHOT package.  
The surface brightness of host galaxies at the locations of both SNe 
was low and did not affect the measurements,
so the subtraction of galaxy background was not necessary. 
The image of SN 2016coi with local standards is shown in Fig. 1, 
the magnitudes of the stars were taken from Kumar et al. (2018).
The results of our photometry are reported in Table 1, and the 
light curves are shown 
in Fig. 2., where we also plotted the data from Kumar et al. (2018),
Prentice et al. (2018), and Terreran et al. (2019).


\begin{figure}
\includegraphics[width=\columnwidth]{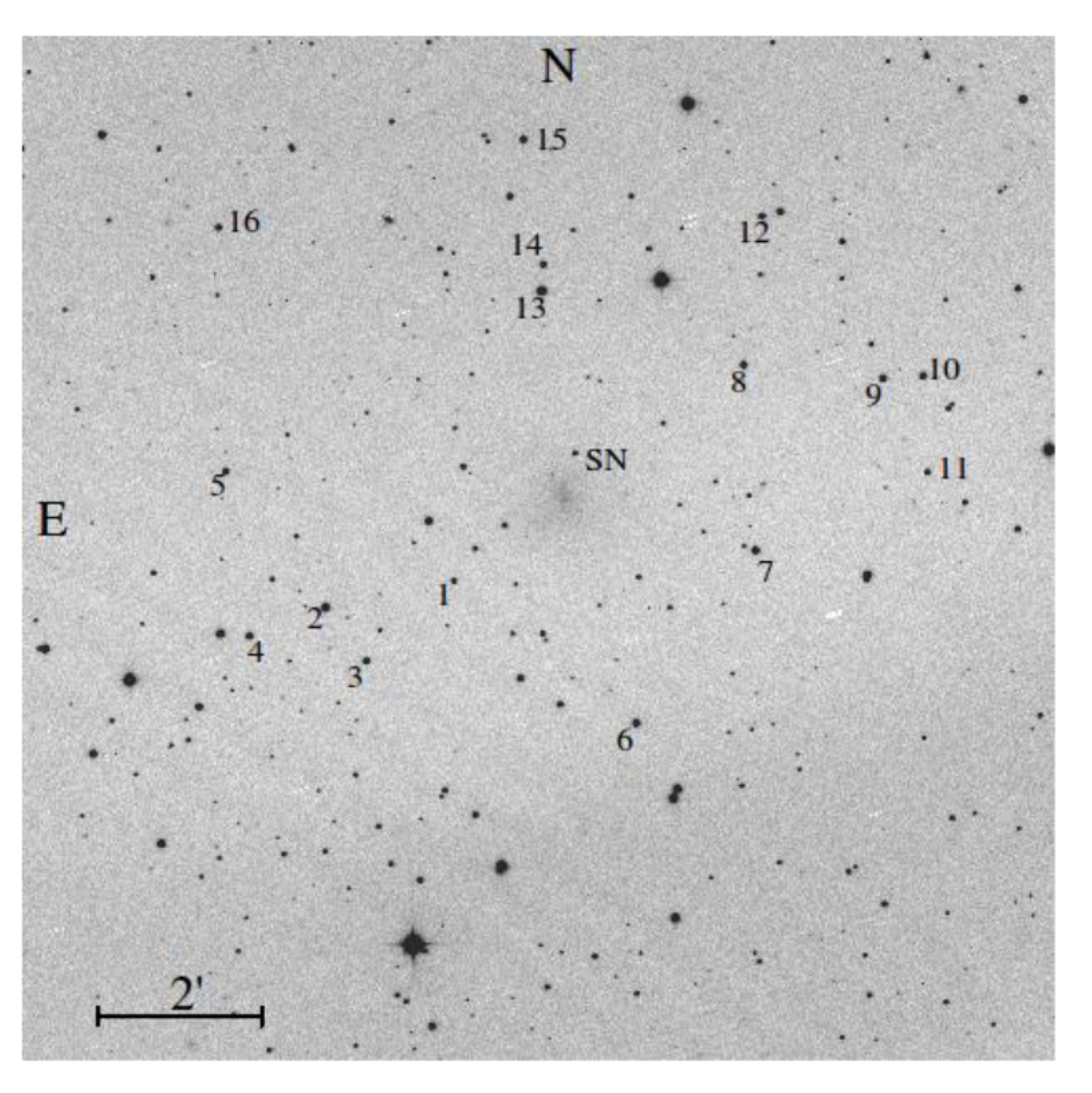}
\caption{SN 2016coi and local standard 
stars from Kumar et al. (2018). The image was obtained with the C60 telescope
in the $R$-band}
\end{figure}


\begin{figure}
\includegraphics[width=\columnwidth]{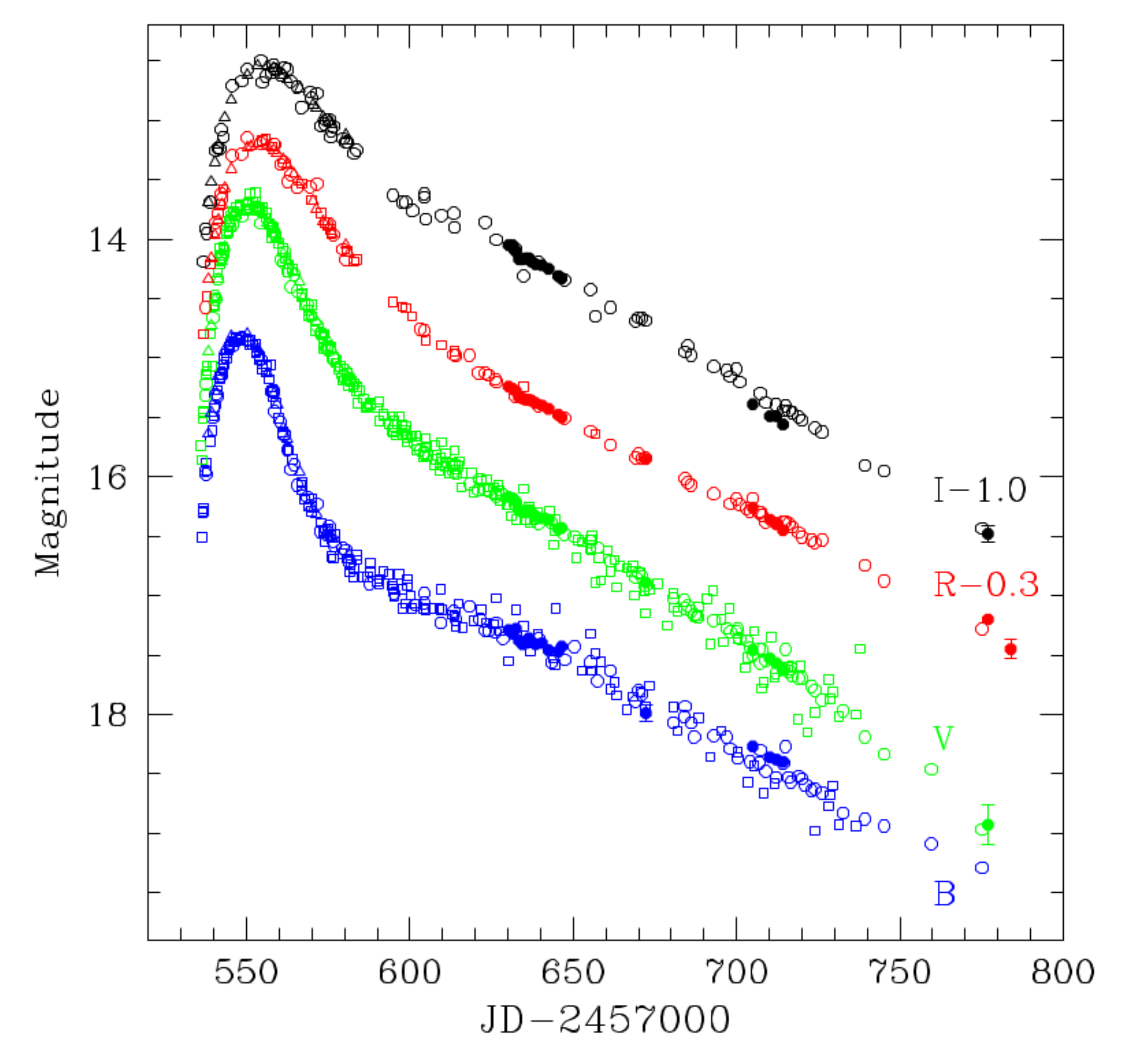}
\caption{The $BVRI$ light curves of SN 2016coi.
Dots show our data; triangles, circles, and squares present the data from
Kumar et al. (2018), Prentice et al. (2018), Terreran et al. (2019)}
\end{figure}

The image of SN 2018ebt is shown in Fig. 3,  
the magnitudes of local standards
are presented in Table 2. The $gri$ magnitudes of local standard stars were 
obtained from the Pan-STARRS 
database\footnote{https://catalogs.mast.stsci.edu/panstarrs/} 
and transformed to 
$BVRI$ magnitudes using relations by Kostov and Bonev (2018).
The errors of $gri$ magnitudes are less than 0.01 mag, and we 
used the dispersion of transformation equations as estimates of the errors
of $BVRI$ magnitudes.

\begin{table}
\caption{Observations of SN 2016coi}\vskip2mm
\begin{tabular}{rccccccccl}
\hline
JD 2457000+ & $B$ & $\sigma_B$ & $V$ & $\sigma_V$ & $R$ & $\sigma_R$ &
$I$ & $\sigma_I$ & Tel.\\
\hline
630.39&  17.29& 0.03 & 16.17& 0.02&  15.54& 0.02&  15.05& 0.02& C60 \\
631.53&  17.32& 0.04 & 16.18& 0.02&  15.56& 0.02&  15.05& 0.02& C60 \\
632.51&  17.28& 0.03 & 16.21& 0.02&  15.58& 0.02&  15.10& 0.02& C60 \\
633.53&  17.38& 0.03 & 16.28& 0.02&  15.63& 0.02&  15.17& 0.02& C60 \\
634.42&  17.41& 0.03 & 16.30& 0.02&  15.64& 0.02&  15.17& 0.02& C60 \\
635.46&  17.40& 0.03 & 16.29& 0.02&  15.65& 0.02&  15.16& 0.02& C60 \\
636.48&  17.36& 0.03 & 16.28& 0.02&  15.65& 0.02&  15.16& 0.02& C60 \\
637.39&  17.39& 0.03 & 16.30& 0.02&  15.66& 0.02&  15.19& 0.02& C60 \\
638.40&  17.41& 0.03 & 16.33& 0.02&  15.68& 0.02&  15.21& 0.02& C60 \\
640.49&  17.40& 0.03 & 16.35& 0.02&  15.70& 0.02&  15.22& 0.02& C60 \\
642.48&  17.46& 0.03 & 16.36& 0.02&  15.73& 0.02&  15.25& 0.02& C60 \\
645.52&  17.47& 0.03 & 16.44& 0.02&  15.78& 0.02&  15.31& 0.02& C60 \\
646.40&  17.43& 0.03 & 16.43& 0.02&  15.80& 0.02&  15.33& 0.02& C60 \\
672.22&  17.99& 0.07 & 16.89& 0.04&  16.15& 0.02&       &     & M70 \\
705.17&  18.27& 0.03 & 17.46& 0.02&  16.56& 0.02&  16.39& 0.02& C60 \\ 
710.22&  18.36& 0.04 & 17.53& 0.03&  16.66& 0.03&  16.49& 0.03& C60 \\
712.21&  18.38& 0.03 & 17.57& 0.02&  16.70& 0.02&  16.49& 0.02& C60 \\
714.27&  18.40& 0.03 & 17.63& 0.02&  16.75& 0.02&  16.56& 0.03& C60 \\
777.16&       &      & 18.93& 0.17&  17.50& 0.03&  17.48& 0.07& S100\\
784.19&       &      &      &     &  17.75& 0.08&       &     & S100\\
\hline
\end{tabular}
\end{table}

\begin{table}
\caption{Magnitudes of local standard stars for SN 2018ebt}\vskip2mm
\begin{tabular}{lcccccccc}
\hline
Star & $B$ & $\sigma_B$ & $V$ & $\sigma_V$ & $R$ & 
$\sigma_R$ & $I$ & $\sigma_I$\\
\hline
1 & 13.72& 0.05&  12.98& 0.03 & 12.53& 0.04 & 12.09& 0.05 \\
2 & 14.88& 0.05&  14.05& 0.03 & 13.56& 0.04 & 13.07& 0.05 \\
3 & 16.72& 0.05&  15.93& 0.03 & 15.45& 0.04 & 14.99& 0.05 \\
4 & 15.81& 0.05&  14.75& 0.03 & 14.09& 0.04 & 13.50& 0.05 \\
5 & 17.35& 0.05&  16.51& 0.03 & 15.99& 0.04 & 15.47& 0.05 \\
6 & 15.77& 0.05&  14.96& 0.03 & 14.46& 0.04 & 13.97& 0.05 \\
7 & 15.68& 0.05&  14.73& 0.03 & 14.16& 0.04 & 13.65& 0.05 \\
8 & 17.74& 0.06&  16.76& 0.03 & 16.17& 0.04 & 15.59& 0.05 \\
9 & 14.62& 0.05&  13.89& 0.03 & 13.45& 0.04 & 13.04& 0.05 \\
10& 15.86& 0.05&  14.86& 0.03 & 14.25& 0.04 & 13.70& 0.05 \\
11& 15.93& 0.05&  14.87& 0.03 & 14.23& 0.04 & 13.70& 0.05 \\
12& 14.72& 0.05&  14.08& 0.03 & 13.69& 0.04 & 13.28& 0.05 \\
\hline
\end{tabular}
\end{table}

\begin{table}
\caption{Observations of SN 2018ebt}\vskip2mm
\begin{tabular}{rccccccccl}
\hline
JD 2458000+ & $B$ & $\sigma_B$ & $V$ & $\sigma_V$ & $R$ & $\sigma_R$ &
$I$ & $\sigma_I$ & Tel.\\
\hline
331.38&  16.46& 0.06&  15.25& 0.04&  14.90& 0.04 & 14.48& 0.06 & M70 \\ 
334.34&  16.68& 0.06&  15.35& 0.04&  14.98& 0.04 & 14.55& 0.06 & M70 \\
341.28&  17.14& 0.06&  15.69& 0.03&  15.21& 0.04 & 14.69& 0.05 & M70 \\
343.33&  17.34& 0.07&  15.90& 0.04&  15.38& 0.04 & 14.80& 0.05 & M70 \\
345.26&  17.54& 0.08&  16.02& 0.04&  15.50& 0.05 & 14.90& 0.05 & M70 \\
348.32&  17.74& 0.07&  16.25& 0.03&  15.67& 0.04 & 15.04& 0.05 & M70 \\
348.55&  17.70& 0.07&       &     &  15.71& 0.05 & 15.12& 0.07 & S100\\
354.31&  18.14& 0.07&  16.63& 0.03&  16.04& 0.04 & 15.38& 0.05 & M70 \\
355.26&  18.12& 0.07&  16.66& 0.03&  16.07& 0.04 & 15.42& 0.05 & M70 \\
358.28&  18.43& 0.08&  16.86& 0.04&  16.27& 0.05 & 15.58& 0.06 & M70 \\
359.46&  18.39& 0.09&  16.92& 0.04&  16.32& 0.04 & 15.61& 0.05 & C60 \\
360.39&  18.34& 0.06&  16.97& 0.04&  16.34& 0.04 & 15.67& 0.05 & C60 \\
360.51&  18.37& 0.07&  16.95& 0.04&  16.35& 0.05 & 15.68& 0.06 & S60 \\
361.44&  18.35& 0.06&  17.00& 0.03&  16.39& 0.04 & 15.69& 0.05 & C60 \\
362.42&  18.37& 0.06&  17.04& 0.03&  16.41& 0.05 & 15.70& 0.06 & C60 \\
368.28&  18.53& 0.11&  17.11& 0.06&  16.59& 0.06 & 15.88& 0.05 & M70 \\
369.35&  18.53& 0.06&  17.23& 0.03&  16.64& 0.05 & 15.93& 0.05 & C60 \\
371.41&  18.51& 0.06&  17.23& 0.04&  16.67& 0.05 & 15.93& 0.06 & C60 \\
374.52&  18.54& 0.07&  17.32& 0.03&  16.75& 0.04 & 16.02& 0.05 & C60 \\
378.51&  18.75& 0.10&  17.40& 0.04&  16.85& 0.05 & 16.08& 0.06 & C60 \\
379.40&  18.72& 0.07&  17.45& 0.03&  16.88& 0.04 & 16.17& 0.05 & C60 \\
380.46&  18.75& 0.07&  17.47& 0.04&  16.92& 0.05 & 16.19& 0.06 & C60 \\
393.23&  18.77& 0.08&  17.67& 0.04&  17.17& 0.05 & 16.52& 0.06 & M70 \\
403.21&  18.95& 0.08&  17.84& 0.04&  17.30& 0.05 & 16.73& 0.05 & M70 \\
404.17&  19.03& 0.09&  17.90& 0.04&  17.29& 0.05 & 16.75& 0.06 & M70 \\
407.16&  19.22& 0.10&  17.95& 0.04&  17.47& 0.06 & 16.84& 0.06 & M70 \\
408.18&  19.11& 0.10&  17.91& 0.05&  17.43& 0.05 & 16.82& 0.06 & M70 \\
410.16&  19.05& 0.13&  17.99& 0.06&  17.41& 0.06 & 16.84& 0.06 & M70 \\
414.17&  19.14& 0.15&  18.15& 0.07&  17.54& 0.06 & 17.00& 0.08 & M70 \\
434.29&  19.54& 0.07&  18.45& 0.04&  17.92& 0.05 & 17.23& 0.05 & C60 \\
437.26&  19.56& 0.09&  18.51& 0.04&  17.99& 0.05 & 17.31& 0.05 & C60 \\
444.37&       &     &  18.69& 0.11&  18.08& 0.07 & 17.32& 0.07 & C60 \\
445.19&       &     &  18.74& 0.17&  18.10& 0.11 & 17.76& 0.09 & C50 \\
446.18&  19.61& 0.07&  18.59& 0.04&  18.13& 0.05 & 17.53& 0.06 & C60 \\
448.20&  19.46& 0.14&  18.74& 0.10&  18.25& 0.10 & 17.63& 0.13 & C60 \\
\hline
\end{tabular}
\end{table}

\begin{figure}
\includegraphics[width=\columnwidth]{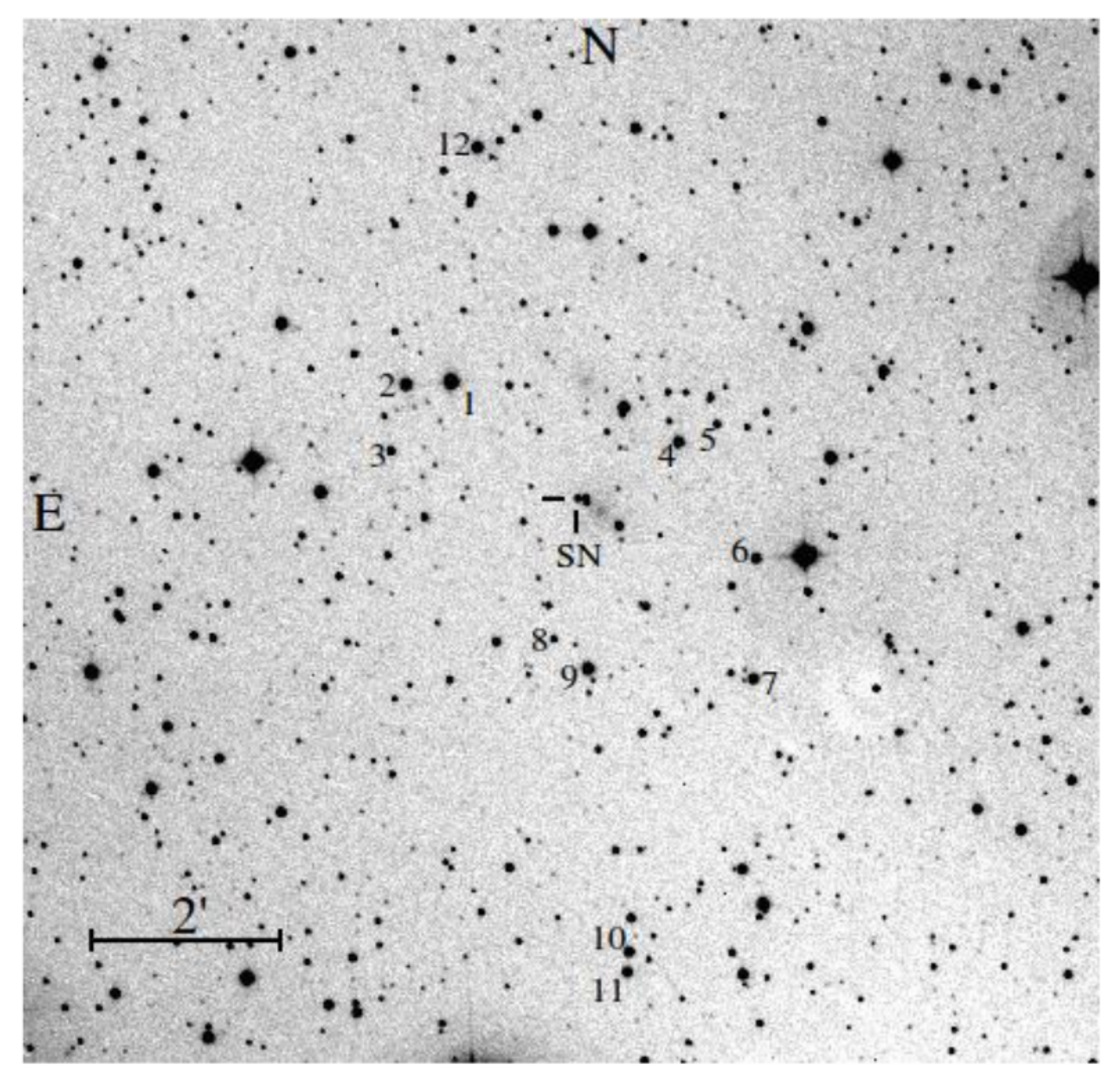}
\caption{SN 2018ebt and local standard 
stars. The image was obtained with the C60 telescope
in the $R$-band}
\end{figure}

\begin{figure}
\includegraphics[width=\columnwidth]{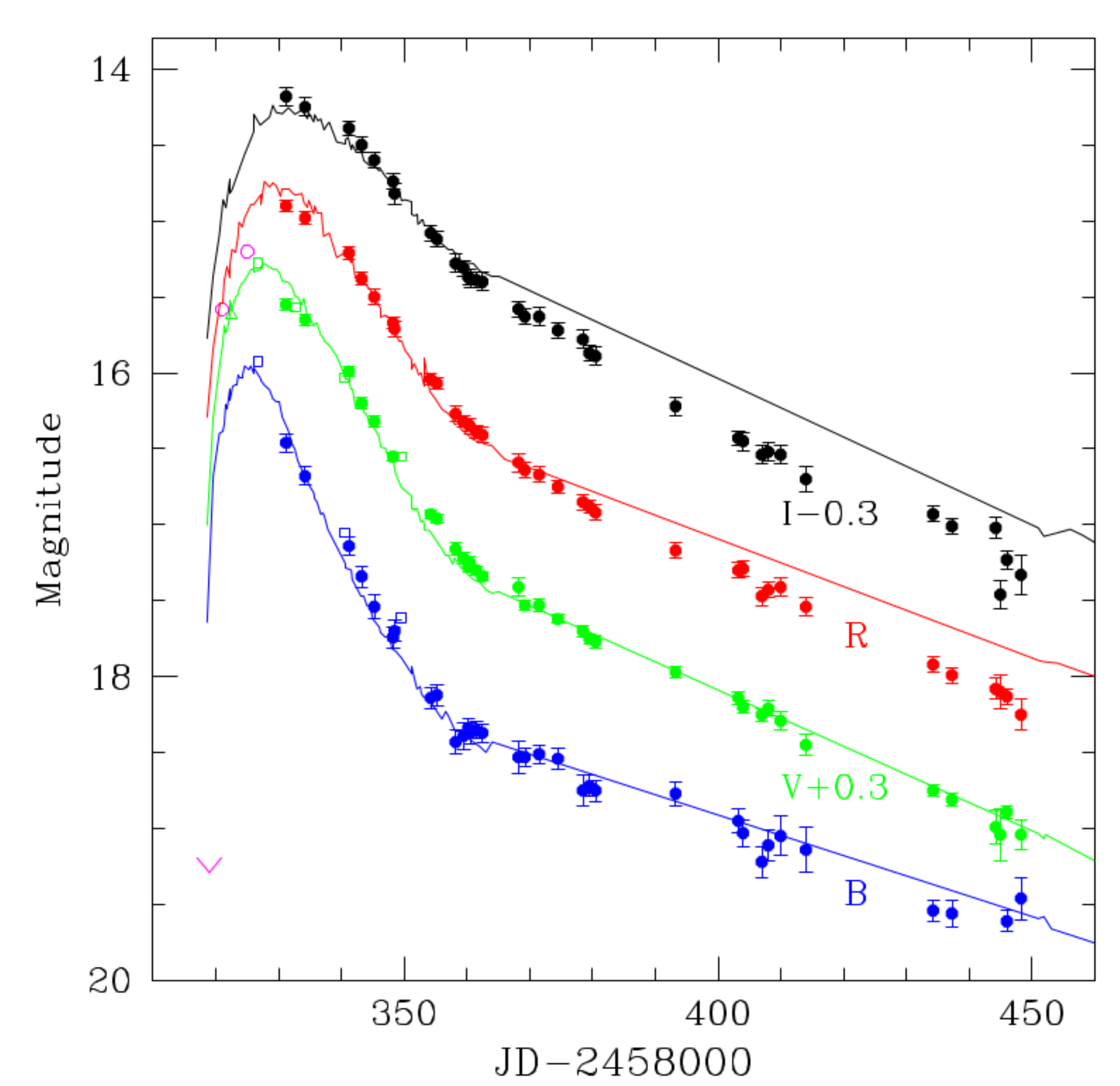}
\caption{The $BVRI$ light curves of SN 2018ebt.
Dots show our data, squares are for the $B,V$ magniudes from the OSC
($B$ magnitudes shifted by 0.35 mag), magenta circles and upper limit mark
show the data from ATLAS. The lines are the light curves of SN 2002ap.}
\end{figure}

\begin{figure}
\includegraphics[width=\columnwidth]{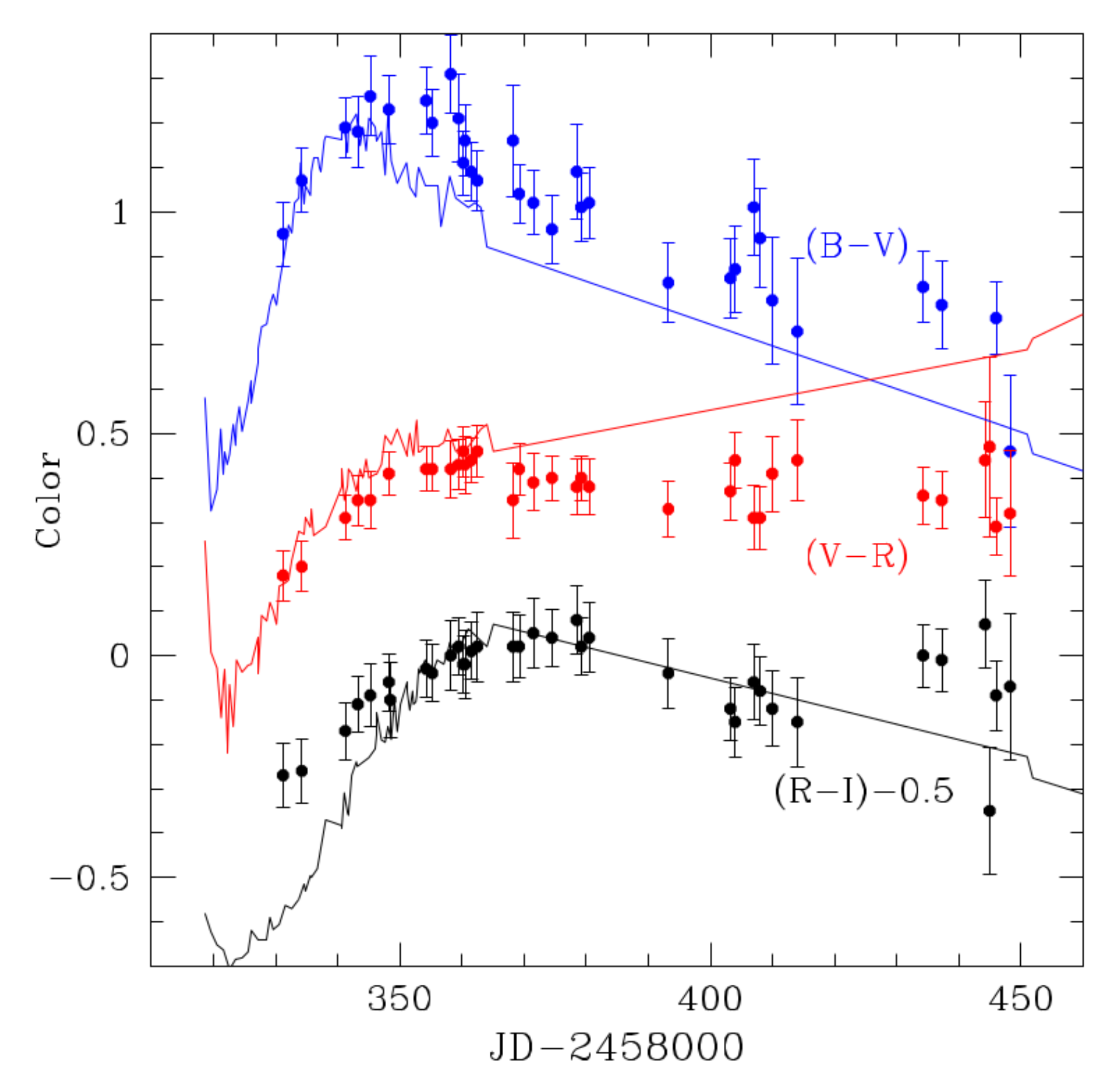}
\caption{The color curves of SN 2018ebt.
Dots show our data, lines are the color curves of SN 2002ap.}
\end{figure}

The $BVRI$ magnitudes of SN 2018ebt are reported in Table 3, and the 
light curves are shown
in Fig. 4. We plotted the data from ATLAS and the $BV$ magnitudes 
reported by the Open Supernova Catalogue (OSC)\footnote{https://sne.space/}.
The $V$-band magnitudes from the OSC showed good agreement with our data,
while the $B$-band magnitudes were significantly brighter. We shifted these
$B$ magnitudes by 0.35 mag for consistency with our data.

\bigskip
\bigskip
\PZsubtitle{\bf Results and conclusions}

{\bf SN 2016coi}.
The light curves shown in Fig. 2 appear typical for type Ib/c SNe. The 
magnitudes from the three major data sets (Kumar et al., 2018; 
Prentice et al.,
2018; Terreran et al., 2019) are in good agreement, although the scatter of
the $B,V$ magnitudes from Terreran et al. (2019) is significantly larger
than of those from the other sources. Our data agree with 
these sets and are useful for studying
the behavior of SN luminosity at late stages. There is a change of 
slope on the tail at about 100 days past maximum, which is evident in the
$B$-band light curve and may be noticed also in the $V$-band; this fact
was not reported in previous studies. The rate of decline in the $B$-band
in the time interval JD 2457590-650 (in mag/day) is 0.0099, while that in 
the 
JD 2457650-780 time interval is 0.0149. For the $V$-band, the 
rates are 0.0158 for
JD 2457590-680 and 0.0211 for JD 2457680-780. In the $R$ and $I$ bands 
the linear decline can be approximated by a single rate, respectively:
0.0141 and 0.0156.  

\medskip
{\bf SN 2018ebt}.
We compared the 
light curves of SN 2018ebt to those for a number of well-studied SNe 
Ic and Ic-BL and found the best match with SN Ic-BL 2002ap (Foley et al., 2003,
Yoshii et al., 2003). The fit of the $B$- and $V$-band light curves is very 
good, but for the $R$ and $I$ bands, the brightness on the linear tail declined 
faster in the case of SN 2018ebt. Using the fit, we can derive dates and 
magnitudes 
of maximum light for SN 2018ebt: $B_{max}=16.00; V_{max}=14.98;
R_{max}=14.72; I_{max}=14.56$. The maximum in the $B$-band was reached on
JD 2458325, and in the other bands about 2-4 days later.
The linear tail of the light curves started at about JD 2458365, 
the rates of brightness decline on the tail 
in the $B,V,R,I$ bands are respectively: 0.0129, 0.0191, 0.0199, 0.0217
mag/day.  

The color curves of SN 2018ebt and SN 2002ap are compared on Fig. 6
after correction for extinction.
The Galactic reddening $E(B-V)=0.26$ mag was accepted for 
SN 2018ebt\footnote{http://ned.ipac.caltech.edu}, and
for SN 2002ap we took $E(B-V)=0.08$ mag (Foley et al., 2003). 
The comparison allows us to conclude that extinction in the host galaxy is
negligible for SN 2018ebt. The shape of the color curves is similar for 
both SNe, although some differences are evident. The $V-R$ color of 
SN 2002ap is redder on the tail, and for the $R-I$ color, the maximum
difference is at stages closer to maximum light.

The redshift of the parent galaxy is unknown, and the redshift estimates  
based on the spectra of the SN are uncertain. The smallest value of
redshift, $z$=0.005, reported by Masayuki Yamanaka via TNS corresponds to
the absolute magnitude at maximum $M_V=-17.4$ mag, which is close to the
value for SN 2002ap, $M_V=-17.1$ mag (Yoshii et al., 2003). We consider this
estimate of $z$ most probable, but the redshift of the host galaxy is 
needed to derive more reliable luminosity of SN 2018ebt and to compare 
this object to other SNe of similar class.    

{\bf Acknowledgments:}
The authors are grateful to S.Yu.Shugarov and I.M.Volkov,
who carried out some of the observations. 

\references
Dugas, A., et al., 2018, {\it ATel}, No. 11974

Elias-Rosa, N., et al., 2016, {\it ATel}, No. 9090

Filippenko, A.V., 1997, {\it ARA\&A}, {\bf35}, 309

Foley, R.J., et al., 2003 {\it PASP}, {\bf115}, 1220

Galama, T.J., et al., 1998, {\it Nature}, {\bf395}, 670

Holoinen, T. W.-S., et al., 2016, {\it ATel}, No. 9086

Kostov, A., Bonev, T., 2018, {\it Bulgarian Astron. J.}, {\bf28}, 3

Kumar, B., et al., 2018, {\it MNRAS}, {\bf473}, 3776

Moran, S., et al., 2018, {\it ATel}, No. 11891

Prentice, S. J., et al., 2018, {\it MNRAS}, {\bf478}, 4162

Terreran, G., et al., 2019, {\it ApJ}, {\bf883}, 147

Valenti, S., et al., 2008, {\it MNRAS}, {\bf383}, 1485

Yamanaka, M., et al., 2017, {\it ApJ}, {\bf837}, 1

Yoshii, Y., et al., 2003, {\it ApJ}, {\bf592}, 467
\endreferences
\end{document}